\documentclass[10pt]{article}
\usepackage{amsmath,amssymb,amsfonts}
\usepackage{units,hyperref,fancyhdr}
\usepackage{bbold,mathpazo}
\usepackage{graphicx}
\usepackage{dcolumn}
\usepackage{bm,bbm,mathtools,esvect}
\usepackage{slashed}
\usepackage{fullpage}
\title{\bf QUANTUM SELF-INTERACTION WITHIN AN INFINITELY DEEP CAVITY \\  \vspace{5mm} }
\author{{\sc SERGIO GIARDINO\footnote{\sc sergio.giardino@ufrgs.br}}\\
\\
\small \it Departamento de Matem\'atica Pura e Aplicada \\
\small \it Universidade Federal do Rio Grande do Sul (UFRGS)\\
\small \it Caixa Postal 15080, 91501-970  Porto Alegre RS \\
\small \it Brazil}

\begin{document}
\date{}
\maketitle

\begin{abstract}
\noindent One examines the infinitely deep quantum cavity, also known as the quantum infinite square well, within the framework of the real Hilbert space.  The solutions are considered in terms of complex wave functions, and also in terms of quaternionic wave functions. The complex results reproduce the usual achievements established in the complex Hilbert space, but also extend them to non-stationary solutions, as well as to distorted stationary solutions, different energy spectra, and dislocated observed position. The quaternionic cases further admit the incidence of self-interaction, something that cannot be observed in complex solutions. Therefore, both the complex and quaternionic solutions are more general than previous cases, thus opening the way to further one-dimensional solutions to be researched in the non-relativistic theory.

\vspace{2mm}

\noindent {\bf keywords:} quantum mechanics; formalism; other topics in mathematical methods in physics

\vspace{1mm}

\noindent {\bf pacs numbers:} 03.65.-w; 03.65.Ca; 02.90.+p. 
\end{abstract}

\vspace{1cm}

\hrule
{\parskip - 0.3mm \footnotesize{\tableofcontents}}
\vspace{1cm}
\hrule

\pagebreak

\section{INTRODUCTION\label{I}}%

Theoretical self-interacting quantum particles \cite{Giardino:2024tvp} have been recently obtained within the ambit of quaternionic quantum mechanics ($\mathbbm H$QM). In the current article, one verifies the self-interaction feature also to be observed within the simple and important physical model of the infinitely deep cavity, also known either as the infinite square well, or as the particle in a box. The solution of the infinite cavity in $\mathbbm H$QM was recently solved in a restricted form \cite{Giardino:2020cee}, but the solutions presented here consider the problem in terms of higher mathematical generality. From a physical standpoint, one notices the self-interacting solutions as not viable within the usual formulation of quantum mechanics in the complex Hilbert space, but are perfectly possible within the more general theory of quaternionic wave funtions within the real Hilbert space \cite{Giardino:2018rhs}. Consequently, this article discusses the presence of self-interaction within particles confined in such cavities on the one hand, while it discusses the generalization of quantum mechanics using quaternions, on the other hand, and also solves the totally general quaternionic cavity for the first time. This generalization of quantum mechanics is done replacing the usual complex numbers ($\mathbbm C$) with quaternionic numbers ($\mathbbm H$), but the mathematical theory of quaternions can be found in several sources \cite{Ward:1997qcn,Morais:2014rqc,Ebbinghaus:1990zah} and will not be considered here in detail. 

Nevertheless, it is important to mention the strong relation between the generalization of quantum mechanics in terms of quaternions, and the character of the Hilbert space. Quantum quaternionic applications can be qualified in terms of the complex, quaternionic, or real character of the Hilbert space. The conventional framework of quantum theory within the complex Hilbert space, here named $\mathbbm C$QM, admits  quaternions as elements of mathematical methods applied to numerous quantum problems, such as quantum states \cite{Steinberg:2020xvf}, wave equations  \cite{Arbab:2010kr,Graydon:2013sra,Sapa:2020dqm,Rawat:2022xsj}, quantum masses \cite{Arbab:2022cpe}, fermions \cite{Cahay:2019bqp,Cahay:2019pse},  monopoles \cite{Soloviev:2016qsx},  and several other things. Conversely, the quaternionic Hilbert space establishes a generalized quantum theory using anti-hermitian Hamiltonian operators in the wave equation, naming anti-hermitian this theory. The large amount of research within anti-hermitian $\mathbbm H$QM occupies a book  by Stephen Adler \cite{Adler:1995qqm}, but this fact does not hide the drawbacks of anti-hermitian $\mathbbm H$QM, such as the ill-defined classical limit ({\it c.f.} sec. 4.4 of \cite{Adler:1995qqm}), which implies the Ehrenfest theorem not to hold. Additional disadvantages concern their involved formulation, where simple solutions are currently difficult to obtain, as well as to interpret. Nevertheless, various research works have been conducted, including scattering states \cite{DeLeo:2015hza,Sobhani:2016qdp,Hassanabadi:2017wrt,Hassanabadi:2017jiz,Procopio:2017vwa,Sobhani:2017yee}, potentials and operators \cite{Ducati:2001qo,Nishi:2002qd,Madureira:2006qps,DeLeo:2013xfa}, wave packets \cite{Ducati:2007wp}, confined states \cite{DeLeo:2005bs,Giardino:2015iia}, perturbation models \cite{DeLeo:2019bcw}, fractional time \cite{Chamizo:2024wzl} and quantum computation \cite{Dai:2023xxh,Nyirahafashimana:2025rho}. 

As the last alternative to quantum quaternionic applications, real Hilbert space $\mathbbm H$QM \cite{Giardino:2018rhs} overcomes the difficulties of the anti-hermitian approach, comprehending well-defined classical limits \cite{Giardino:2018lem}, and simple quantum solutions to models like the geometric phase \cite{Giardino:2016xap}, the autonomous particle \cite{Giardino:2017yke,Giardino:2017pqq,Giardino:2024tvp}, the Virial theorem \cite{Giardino:2019xwm,Giardino:2025bym}, the step potential \cite{Giardino:2020cee}, the harmonic oscillator, the elastic scattering \cite{Giardino:2020ztf,Hasan:2019ipt}, the harmonic oscillator \cite{Giardino:2021ofo}, spin \cite{Giardino:2023spz}, and the generalized wave equation \cite{Giardino:2023uzp}. Relativistic theories were also entertained, embracing the Klein-Gordon equation \cite{Giardino:2021lov}, the Dirac equation \cite{Giardino:2021mjj}, scalar fields \cite{Giardino:2022kxk}, and Dirac fields \cite{Giardino:2022gqn}. 

The infinitely deep cavity can be considered as a particular example of a rectangular potential, and besides their interest as a means to understand the quaternionic theory, rectangular potentials have their own interest, and innovative research is currently conducted. Recent examples comprise quantum cavities \cite{Vincenzo:2018kjv,Giordano:2022qec,Albrecht:2022sdd,Cohen:2022pdh,Vedral:2024iab,Kim:2024bcx}, the double well \cite{Sasaki:2022aby,Fassari:2023twe}, circular wells \cite{Levi:2022ipj,ghosh:2021nrs}, oscillators \cite{Salamu:2023bme}, topological cavities \cite{deOliveira:2023nwv}, quantum time \cite{vega:2021qqt} supersymmetry \cite{Gutierrez:2018uyz}, phase space \cite{Rojo:2020iif}, fractional theory \cite{Wei:2015thi}, entanglement \cite{Zhang:2025jpp}, hermiticity \cite{Burgess:2016lal,Liu:2025ius}, and a whole review subsumes recent research concerning this theme \cite{Belloni:2014isq}. In closing, the results found in this article confirm the suitability of $\mathbbm H$QM in the real Hilbert space as a viable theoretical description for quantum phenomena, and future research will hopefully further verify this viability.

\section{THE COMPLEX CASE}
 
The infinitely deep cavity potential is one of the simplest solutions of quantum mechanics. In this section, one considers it using the real Hilbert space approach using a complex wave function that will serve as reference for the quaternionic solutions of the next section, as well as to point out the differences to the usual $\mathbbm C$QM result. After noticing the autonomous particle wave function in the real Hilbert space \cite{Giardino:2024tvp}  as the main reference to this section, one introduces the potential
\begin{equation}\label{ci05}
\mathcal V(x)=\left\{
\begin{array}{ll}
V      & \qquad \mbox{if}\qquad x\in\left[-\nicefrac{\ell}{2},\,\nicefrac{\ell}{2}\right] \\ \\
\infty & \qquad \mbox{if} \qquad x\in\left(-\infty,\,-\nicefrac{\ell}{2}\right)\cup \left(\nicefrac{\ell}{2},\,\infty\right).
\end{array}
\right.
\end{equation}
where $\ell$ is the length of the cavity, and $V$ is a complex constant such as
\begin{equation}
 V=V_0+V_1 i,
\end{equation}
whose real constant components are of course $V_0$ and $V_1$.  The solutions to be considered in this paper exclude the region outside of the well, although it is possible to include them using operator techniques, as it is found in \cite{Carreau:1990lal}.  Developing Setting $V=0$ recovers the usual infinite square cavity potential entertained in within standard $\mathbbm C$QM. Associated to a quantum particle of mass $m$ trapped within this cavity, there is a complex wave function $\psi$ that complies with the ordinary Schr\"odinger equation,
\begin{equation}\label{ci01}
 i\hbar\frac{\partial \psi}{\partial t}=\left(-\frac{\hbar^2}{2m}\nabla^2+\mathcal V\right)\psi.
\end{equation}
In the one-dimensional region where the particle is allowed to move, the wave function comprises
\begin{equation}\label{ci02}
 \psi(x,\,t)=\phi\big( x\big)\exp\left[-\frac{E}{\hbar}t\right]
\end{equation}
whereby the energy parameter $E$ is complex. The space dependent function $\phi(x)$ comprehends
\begin{equation}\label{ci34}
 \phi(x)=A\exp\left[Kx\right]+B\exp\left[-Kx\right],
\end{equation}
and the linear momentum parameter $K$ is also complex, so that
\begin{equation}\label{ci22}
 E=E_0+E_1i,\qquad\qquad K=K_0+K_1i,
\end{equation}
remembering $E_0,\,E_1,\,K_0$ and $K_1$ to be real constants, and the amplitudes $A$ and $B$ to be complex numbers. The wave function (\ref{ci02}) is purposely written in terms of explicit components, representing two independent autonomous particles that move on opposite directions, although it of course represents one single particle trapped in the cavity.

The substitution of the wave function (\ref{ci02}) in the wave equation (\ref{ci01}) generates two real constraints
\begin{equation}\label{ci04}
 K_0^2-K_1^2=\frac{2m}{\hbar^2}\Big(V_0-E_1\Big),\qquad \mbox{and}\qquad 2 K_0 K_1=\frac{2m}{\hbar^2}\Big(V_1+E_0\Big),
\end{equation}
where the first constraint is related to the conservation of the energy, and the second relates itself to the conservation of the probability, as will be later discussed. Therefore, (\ref{ci04}) consequently generate
\begin{equation}\label{ci26}
 K_0^2=\frac{m}{\hbar^2}\left(V_0-E_1+\sqrt{(E_1-V_0)^2+(V_1+E_0)^2}\,\right)
\end{equation}
and 
\begin{equation}\label{ci27}
 K_1^2=\frac{m}{\hbar^2}\left(E_1-V_0+\sqrt{(E_1-V_0)^2+(V_1+E_0)^2}\,\right).
\end{equation}
Nevertheless, steady state requirements may be ascertained, whereby
\begin{equation}\label{ci18}
V_1+E_0=0\qquad \mbox{and}\qquad V_0<E_1, \qquad \mbox{imply that}\qquad K_0=0,\qquad \mbox{and}\qquad K_1\neq 0,
\end{equation}
thus determining a pure oscillatory stationary motion. Inversely, the totally non-oscillatory behaviour demands that
\begin{equation}\label{ci19}
V_1+E_0=0\qquad \mbox{and}\qquad E_1<V_0, \qquad \mbox{imply that}\qquad K_0\neq 0,\qquad \mbox{and}\qquad K_1= 0.
\end{equation}
Evidently, the most general alternative occurs whenever that $V_1+E_0\neq 0$ imposes non-zero real factors to $K$, and allows non stationary oscillatory motion along the spatial parameter $x$.

The above solution is of course valid independent of the version of quantum mechanics, because it depends on the Schr\"odinger equation only. However, the definition of the expectation value of an operator $\widehat{\mathcal O}$ in the real Hilbert space \cite{Giardino:2018rhs}, specifically
\begin{equation}\label{ci03}
\big\langle\widehat{\mathcal O}\big\rangle =\frac{1}{2}\int d\bm x\left[\Psi^\dagger \widehat{\mathcal O}\Psi+\Big(\widehat{\mathcal O}\Psi\Big)^\dagger\Psi\right],
\end{equation}
differs decisively from the ordinary complex Hilbert space approach, remembering $\Psi^\dagger$ to be the adjoint of the wave function $\Psi$. Besides (\ref{ci03}) is evaluated over real numbers, operator  $\widehat{\mathcal O}$ is arbitrary, and not compulsorily hermitian, therefore, the real Hilbert space approach is more general because it is submitted to fewer constraints. Following this formalism, the required definitions of the energy, and the linear momentum operators,
\begin{equation}
 \widehat E=i\hbar\frac{\partial}{\partial t},\qquad  \mbox{and}\qquad
 \widehat p=-i\hbar\frac{\partial}{\partial x},
\end{equation}
as well as the wave function (\ref{ci02}), together produce
\begin{eqnarray}
\nonumber&& \left\langle \widehat E\right\rangle=E_1\int\!\!\rho \,dx,\\
\nonumber && \Big\langle \widehat{\bm p}\Big\rangle=\hbar K_1 \int\!\!\rho\, dx,\\
\label{ci08}&& \left\langle \widehat{p^2}\right\rangle=\hbar^2\Big(K^2_1-K_0^2\Big) \int\!\!\rho\, dx,\\
\nonumber&&\left\langle \widehat V\right\rangle=V_0\int\!\!\rho\, dx,
\end{eqnarray}
recalling that the probability density $\rho$ equals 
\begin{equation}
 \rho=|\psi|^2.
\end{equation}
Using the conservation of the mechanical energy relation
\begin{equation}\label{ci09}
 \left\langle \widehat E\right\rangle=\frac{1}{2m}\left\langle \widehat p^2\Big\rangle+\Big\langle \widehat V\right\rangle,
\end{equation}
one observes the first relation (\ref{ci04}) to be held on both of the exponential functions that compose the wave function (\ref{ci02}-\ref{ci34}) after the integral of the probability density factors out. Relation (\ref{ci09}) also holds in case of negative kinetic energy, a situation observed wherever $K_1^2<K_0^2$, and $E_1<V_0$ in (\ref{ci04}), and that imposes the conservation of the energy to be understood as a local phenomenon. Put in other words, even when the energy is not conserved because of non-zero real components either of the energy $E$, or of the linear momentum $K$, the real exponential function factors out and (\ref{ci04}) holds at every instant of time, and at every point of the space, something that cannot be obtained in $\mathbbm C$QM, where non-stationary processes cannot be understood in terms of the conservation of the energy. 

Moreover, one has to point out the second relation of (\ref{ci04}) as the imaginary part generated after the substitution of the wave function in the Schr\"odinger equation, and it has been demonstrated \cite{Giardino:2024tvp} to be the continuity equation. This constraint is non-trivial in case of non-stationary processes, also observed in case of complex potentials within the usual $\mathbbm C$QM ({\it cf.} \cite{Schiff:1968qmq} Section 20). 

In summary, the real Hilbert space approach provides a general framework, where stationary and non stationary processes can be jointly understood, something inaccessible in terms of the complex Hilbert space approach to quantum mechanics. Therefore, stationary and non-stationary solutions to a particle moving within a infinitely deep cavity can be considered in the sequel. There are several possible boundary conditions to the wave function submitted to (\ref{ci05}), the symmetric one encompasses 
\begin{equation}\label{ci06}
 \psi\big(-\nicefrac{\ell}{2}\big)=\psi\big(\nicefrac{\ell}{2}\big)\qquad \mbox{implying} \qquad \Big(A-B\Big)\Big(1-\exp\big[K\ell\big]\Big)=0,
\end{equation}
whereas the anti-symmetric boundary condition comprises
\begin{equation}\label{ci07}
 \psi\big(-\nicefrac{\ell}{2}\big)=-\psi\big(\nicefrac{\ell}{2}\big)\qquad \mbox{implying} \qquad \Big(A+B\Big)\Big(1+\exp\big[K\ell\big]\Big)=0.
\end{equation}
The boundary conditions respect the bilateral symmetry of the potential because the density of probability of $|\psi|^2$ is equal on both of the sides of the cavity, as it seems reasonable, and thus one merge conditions (\ref{ci06}-\ref{ci07}) within a single condition
\begin{equation}\label{ci43}
\left| \psi\big(-\nicefrac{\ell}{2}\big)\right|^2=\left|\psi\big(\nicefrac{\ell}{2}\big)\right|^2.
\end{equation}
An analysis concerning more general boundary than (\ref{ci43}) has been done within the ambit of the complex Hilbert space \cite{Vincenzo:2018kjv}, and their discussion in terms of the real Hilbert space approach is an exciting direction for future research. One then considers each possible solution.

\paragraph{SOLUTION I} This solution is ordinarily found in textbooks of quantum mechanics. It is interesting to notice that condition (\ref{ci43}) does not require the wave function to be zero on the boundaries, as usually found. Conditions (\ref{ci06}-\ref{ci07}) admit the symmetric and anti-symmetric conditions
\begin{equation}\label{ci10}
 A=B,\qquad \exp\big[K\ell\big]=1 \qquad \mbox{and}\qquad A=-B,\qquad \exp\big[K\ell\big]=-1,
\end{equation}
and both of them require $K_0=0$. Each condition determines a wave function and a quantization condition, which can be arranged together as
\begin{eqnarray}\label{ci37}
 \psi_N(x,\,t)=
 \left\{
 \begin{array}{ll}
  \sqrt{\frac{2}{\ell}}\cos\Big(N'\delta_{NN'}\frac{\pi}{\ell}x\Big)\exp\left[-\frac{E}{\hbar}t\right],\qquad \mbox{if}\qquad N=2n-1\\ 
  \sqrt{\frac{2}{\ell}}\sin\Big(N'\delta_{NN'}\frac{\pi}{\ell}x\Big)\exp\left[-\frac{E}{\hbar}t\right],\qquad \mbox{if}\qquad N=2n
 \end{array}
 \right.
\end{eqnarray}
and $N$ obviously represents a positive non zero integer number. It also holds the orthogonality relation within the time independent component of the wave function, namely
\begin{equation}\label{ci17}
 \big\langle \phi_N,\,\phi_M\big\rangle =\delta_{NM}.
\end{equation}
The  determined real components of the complex parameter $K$, and (\ref{ci04}) inevitable produce
\begin{equation}\label{ci11}
E_1=\frac{\hbar^2N^2\pi^2}{2m\ell^2}+V_0,\qquad \mbox{and}\qquad E_0+V_1=0,
\end{equation}
and consequently the expectation values within the real Hilbert space (\ref{ci08}-\ref{ci09}) generate
\begin{equation}\label{ci13}
 \langle E\rangle =E_1\exp\left[\frac{2V_1}{\hbar}t\right],\qquad \langle x \rangle = \langle p \rangle =0,
\end{equation}
thus recovering the usual $\mathbbm C$QM solution in the case wherever $V_1=0$, as desired. The energy increases or diminishes according to a non zero $V_1$, however keeping (\ref{ci11}) valid,  and then verifying the generality the real Hilbert space formalism. Because of their greater generality, the real Hilbert space formalism deserves to be applied to further solutions of the infinitely deep cavity that are not usually considered within the conventional version of quantum mechanics, and that are entertained in the sequel.

\paragraph{SOLUTION II} In this case, one observes that 
\begin{equation}
 A\neq B\qquad \mbox{and either}\qquad \exp\big[K\ell\big]=1 \qquad \mbox{or}\qquad \exp\big[K\ell\big]=-1,
\end{equation}
also satisfy the conditions (\ref{ci06}-\ref{ci07}), The two quantization conditions emerge can be summarized as
\begin{equation}
 K\ell=iN\pi
\end{equation}
where $N$ is a non-zero integer, and therefore the wave function encompasses
\begin{equation}\label{ci16}
 \psi_N^{II}(x,\,t)=\frac{1}{\sqrt{\ell}}\left(\cos\Theta\exp\left[i\left(\frac{N\pi}{\ell}x+\frac{\Omega}{2}\right)\right]+\sin\Theta\exp\left[-i\left(\frac{N\pi}{\ell}x+\frac{\Omega}{2}\right)\right]\right)\exp\left[-\frac{E}{\hbar}t\right],
\end{equation}
where $\Theta$ and $\Omega$ are a real parameters, and one expects the trigonometric parameters depending on $\Theta$ not to be zero. The wave function satisfies an orthogonality relation like (\ref{ci17}), and one calls $\Theta$ and $\Omega$ the distortion parameters because of the deformation they promote in  wave function (\ref{ci37}) which is a particular case of (\ref{ci16}), reached wherever
\begin{equation}
 \big|\sin\Theta\big|=\big|\cos\Theta\big|=\frac{1}{\sqrt{2}}\qquad \mbox{and}\qquad \Omega=0.
\end{equation}
The physical interpretation is similar to the previous case, where the orthogonality condition (\ref{ci17}) hold as well as the energy properties (\ref{ci11}-\ref{ci13}). However, position and linear momentum expectation values do not recover (\ref{ci13}), and conversely 
\begin{equation}
 \langle x \rangle =\cos N\pi
 \frac{\ell\sin 2\Theta\sin\Omega}{2N\pi}\exp\left[\frac{2V_1}{\hbar}t\right]\qquad \mbox{and}\qquad \langle p \rangle =0.
\end{equation}
These results are somewhat surprising, and  announce quantized position and momentum expectation values within the infinitely deep cavity, and a physically meaningful result requires $V_1\leq 0$. The higher the energy level, the closer expectation value to $\langle x \rangle =0$, and a simple approximation reveals that
\begin{equation}
 -\frac{\ell}{2N\pi}<\langle x\rangle <\frac{\ell}{2N\pi}.
\end{equation}
One cannot determine from these solutions the precise physical role of the distortion parameters $\Theta$ and $\Omega$, but the symmetry breaking promoted by them is a very interesting direction for future research.

\paragraph{SOLUTION III} In this case, the boundary conditions satisfying (\ref{ci06}-\ref{ci07}) are  
\begin{equation}\label{ci14}
 A= B,\qquad A=-B,\qquad \mbox{and}\qquad \exp\big[K\ell\big]\neq\pm 1.
\end{equation}
These solutions do not satisfy quantization conditions, even though satisfy the Schr\"odinger equation, and therefore each solution admits a single quantum state. Single state solutions have already been found within Dirac delta potentials, and therefore cannot be considered a conceptual novelty, although they are normally not found in the infinitely deep cavity. One can consider separately three different cases.

\paragraph{SOLUTION III - THE PROPAGATING CASE} This physical situation corresponds to (\ref{ci18}), where the complex parameter $K$ is pure imaginary, and thus
\begin{equation}
 K_0=0,\qquad \mbox{and}\qquad V_0<E_1.
\end{equation}
The wave function accordingly comprises two possibilities according to (\ref{ci14})
\begin{equation}\label{ci24}
 \psi^{III}(x,\,t)=
 \left\{
 \begin{array}{l}
  \sqrt{\frac{2K_1}{K_1\ell+\sin K_1\ell}}\cos(K_1x)\exp\left[-\frac{E}{\hbar}t\right]\\
  \sqrt{\frac{2K_1}{K_1\ell+\cos K_1\ell}}\sin(K_1x)\exp\left[-\frac{E}{\hbar}t\right],
 \end{array}
\right.
\end{equation}
remembering that $E_1$ and $V_0$ are a free parameters, $E_0=-V_1$, and $K_1$ depends on $E$ and $V$ according to (\ref{ci04}), and the physical expectation values are reproduce (\ref{ci13}), thus remembering a classical oscillation.

\paragraph{SYMMETRIC SOLUTIONS III - THE NON-PROPAGATING CASE} Corresponding to (\ref{ci19}), where the complex parameter $K$ is purely real, so that
\begin{equation}
 K_1=0,\qquad \mbox{and}\qquad E_1<V_0.
\end{equation}
The normalized wave function inevitably comprises
\begin{equation}\label{ci25}
\psi^{III}(x,\,t)=
\left\{
\begin{array}{l}
\sqrt{\frac{2K_1}{K_1\ell+\sinh K_1\ell}}\cosh(K_1x)\exp\left[-\frac{E}{\hbar}t\right]\\
\sqrt{\frac{2K_1}{K_1\ell+\cosh K_1\ell}}\sinh(K_1x)\exp\left[-\frac{E}{\hbar}t\right]
\end{array}
\right.
\end{equation}
It is remarkably that such a solution admits physical expectation values identical to (\ref{ci13}), and also analogous to the propagating case above. The observable difference between the propagating and the non propagating cases is basically the kinetic energy, that is positive in (\ref{ci24}) the propagating case, and negative in (\ref{ci25}).

\paragraph{SOLUTION III - THE COMBINED CASE} In this situation, 
\begin{equation}
 E_0+V_1\neq 0,
\end{equation}
and (\ref{ci26}-\ref{ci27}) demonstrate that $K$ contain a nonzero real component. The wave function therefore assumes
\begin{equation}\label{ci38}
\psi^{III}(x,\,t)=\left\{
\begin{array}{l}
\frac{1}{2\sqrt{\sin\frac{K_1\ell}{2}+\sinh\frac{K_0\ell}{2}}}\Big(e^{Kx}+e^{-Kx}\Big)\exp\left[-\frac{E}{\hbar}t\right]\\
\frac{1}{2\sqrt{\cos\frac{K_1\ell}{2}+\cosh\frac{K_0\ell}{2}}}\Big(e^{Kx}-e^{-Kx}\Big)\exp\left[-\frac{E}{\hbar}t\right].
\end{array}
\right.
 \end{equation}
As in the previous cases, the physical expectation values reproduce (\ref{ci13}), and the kinetic energy can be negative or positive, depending on the relation between the real components of $K$.  

As a final comment, the physically reasonable boundary condition (\ref{ci43}) admits the generalization
\begin{equation}\label{ci44}
\psi\big(-\nicefrac{\ell}{2}\big)=\psi\big(\nicefrac{\ell}{2}\big)\exp[i\omega],
\end{equation}
where $\omega$ is a real quantity, and the space dependent wave function in this case reads
\begin{multline}
\phi_{II}= B\left\{\Bigg(\exp\left[-Q\left(\frac{\ell}{2}-x\right)\right]-\exp\left[Q\left(\frac{\ell}{2}-x\right)\right]\Bigg)\exp\big[i\omega\big]\right.+\\
\left.+\exp\left[-Q\left(\frac{\ell}{2}+x\right)\right]-\exp\left[Q\left(\frac{\ell}{2}+x\right)\right]\right\}.
\end{multline}
However, the position and momentum reproduce (\ref{ci13}), and quantized energies require (\ref{ci44}) to be identically zero. Therefore, this case does not contain physical novelties and does not deserve further discussion. The complex case is then exhausted, and it will be the basis for the quaternionic cases to be considered in the next sections.

\section{QUATERNIONIC CAVITY I}

Before considering the deeply infinite quaternionic cavity, one needs to remember a few basic facts and definitions on quaternions, and on $\mathbbm H$QM in the real Hilbert space, but particularly the autonomous solution \cite{Giardino:2024tvp}. A quaternionic wave function comprises two complex functions, namely $\psi_0$ and $\psi_1$, so that
\begin{equation}\label{ci20}
 \Psi=\psi_0+\psi_1 j,
\end{equation}
where $j$ an quaternionic imaginary unit. One notices the basic elements of quaternion to be found elsewhere \cite{Ward:1997qcn,Morais:2014rqc,Ebbinghaus:1990zah}, nevertheless remembering the anti-commutative relation between imaginary units
\begin{equation}
 ij=-ji,
\end{equation}
which determines that
\begin{equation}
 \Psi i\neq i\Psi.
\end{equation}
Due to this, two alternative wave equations emerge, firstly  
\begin{equation}\label{ci21}
 i\hbar\frac{\partial\Psi}{\partial t}=\widehat{\mathcal H}\Psi,
\end{equation}
while in the remaining possibility the imaginary unit $i$ multiplies the time derivative of the wave function from the right hand side, and this case appears within the next section. The Hamiltonian operator $\widehat{\mathcal H}$, establishes
\begin{equation}
 \widehat{\mathcal H}=-\frac{\hbar^2}{2m}\nabla^2+\mathcal U,
\end{equation}
wherein the quaternionic scalar potential $\mathcal U$ comprises
\begin{equation}\label{ci46}
\mathcal U=U_0+U_1j,
\end{equation}
whose complex components read
\begin{equation}
 U_0=V_0+V_1 i,\qquad\qquad U_1=W_0+W_1 i,
\end{equation}
and evidently   $V_0,\,V_1,\,W_0$ and $W_1$ are real. The quaternionic wave function (\ref{ci20}) and the	 wave equation (\ref{ci21})  generate a  system of two coupled complex equations, definitely
\begin{eqnarray}
\nonumber    && i\hbar\frac{\partial\psi_0}{\partial t}=-\frac{\hbar^2}{2m}\nabla^2\psi_0+U_0\psi_0-U_1\psi_1^\dagger\\
\label{ci28} && i\hbar\frac{\partial\psi_1}{\partial t}=-\frac{\hbar^2}{2m}\nabla^2\psi_1+U_0\psi_1+U_1\psi_0^\dagger,
\end{eqnarray}
wherein to the adjoint of the complex function $\psi$ corresponds $\psi^\dagger$. Blatantly, the complex component $U_1$ of the quaternionic scalar potential $\mathcal U$ promotes a coupling between the complex wave functions $\psi_0$ and $\psi_1$ that can be understood as a self-interaction within the quaternionic wave function. Therefore, one has to separate the self-interacting solutions considered in this paper, and the previously known non-self-interacting solutions \cite{Giardino:2020cee} that can be briefly remembered in the sequel.

\paragraph{NON-SELF-INTERACTING SOLUTIONS} Remarking from (\ref{ci28}) that $U_1=0$ decouples $\psi_0$, and $\psi_1$, and to each component corresponds a proper energy. In this situation, each equation of (\ref{ci28}) generates an independent autonomous particle solution. The simplest case is thus,
\begin{equation}
 \Psi_{nm}=\cos\theta\psi_n+\sin\theta\psi_mj,
\end{equation}
where $\psi_n$ and $\psi_m$ are complex autonomous solutions for to the finitely deep cavity, and $\theta$ is a real parameter. The expectation values for the position and linear momentum recovers the complex case, and one remarks the energy of each orthogonal quantum state to be
\begin{equation}
 \left\langle E_{nm}\right\rangle=\cos^2\theta E_n+\sin^2\theta E_m,
\end{equation}
where $E_n$ and $E_m$ are energetic parameters of each independent case.  The complete discussion of this solution can be found in \cite{Giardino:2020cee}, and will no be further detailed here, because our aim is to consider the coupled $U_1\neq 0$ case.

\paragraph{SELF-INTERACTING AUTONOMOUS PARTICLE} This topic recalls the quaternionic autonomous particle solutions, which will compose the infinitely deep quaternionic cavity. The general solution for a constant potential corresponds to the autonomous particle has been ascertained in \cite{Giardino:2024tvp} to be
\begin{equation}\label{ci23}
 \Psi=\mathcal A\exp\left[K x-\frac{E}{\hbar}t\right],
\end{equation}
where  $K$ and $E$ follow (\ref{ci22}), and the quaternionic amplitude $\mathcal A$ is so that
\begin{equation}\label{ci35}
 \mathcal A=A_0+A_1j,
\end{equation}
where $A_0$ and $A_1$ are complex quantities. Although (\ref{ci23}) is very similar to the complex autonomous case, which is recovered if $A_1=0$, several differences baldly appear. The expectation values deviate from the complex result (\ref{ci08}) because the wave function (\ref{ci23}), within the inner product (\ref{ci03}), generates  
\begin{eqnarray}
\nonumber&& \left\langle \widehat E\right\rangle=E_1\int\!\!\rho \,dx,\\
\nonumber && \Big\langle \widehat{\bm p}\Big\rangle=\hbar K_1 \int\!\!\rho\, dx,\\
\label{dd33} && \left\langle \widehat{p^2}\right\rangle=\hbar^2\Big(K^2_1-K_0^2\Big) \int\!\!\varrho\, dx,\\
\nonumber&&\left\langle \widehat V\right\rangle=V_0\int\!\!\varrho\, dx,
\end{eqnarray}
wherein the probability densities comprise
\begin{equation}
 \rho=\Big(|A_0|^2-|A_1|^2\Big)|\psi|^2,\qquad \mbox{and} \qquad \varrho=\Big(|A_0|^2+|A_1|^2\Big)|\psi|^2,
\end{equation}
and thus recovering the complex instance within the limit of $A_1=0$, as desired. Another fundamental element to be determined is the complex parameter $K$ in terms of the free parameters $E$ and $\mathcal U$. The matrix equation provided by the system (\ref{ci28}) reads
\begin{equation}\label{ci29}
\left[
 \begin{array}{cc}
   U_0+i E & -\, U_1\\
  \overline{ U}_1 & \overline{ U}_0- i\, E
  \end{array}
\right]\left[
\begin{array}{c}
 A_0 \\
 \overline A_1
\end{array}
\right]
=
\frac{\hbar^2K^2}{2m}
\left[
\begin{array}{c}
 A_0 \\
 \overline A_1
\end{array}
\right].
\end{equation}
After establishing
\begin{equation}
 \alpha=\Big(E_0+ V_1\Big)^2- E_1^2+ U_1\overline{U}_1\qquad \mbox{and}\qquad \beta=2 E_1\big( E_0+ V_1\big),
\end{equation}
one accomplishes that
\begin{equation}
 K_0^2=\frac{m}{\hbar^2}\left[V_0\pm\sqrt{\frac{\sqrt{\alpha^2+\beta^2}-\alpha}{2}}+\sqrt{\left(V_0\pm\sqrt{\frac{\sqrt{\alpha^2+\beta^2}-\alpha}{2}}\,\right)^2+\frac{\sqrt{\alpha^2+\beta^2}+\alpha}{2}}\;\right ]
\end{equation}
and  that
\begin{equation}
 K_1^2=\frac{m}{\hbar^2}\left[-V_0\mp\sqrt{\frac{\sqrt{\alpha^2+\beta^2}-\alpha}{2}}+\sqrt{\left(V_0\pm\sqrt{\frac{\sqrt{\alpha^2+\beta^2}-\alpha}{2}}\,\right)^2+\frac{\sqrt{\alpha^2+\beta^2}+\alpha}{2}}\;\right ].
\end{equation}
From the components of $K^2$, the complex components $A_0$ and $A_1$ of the eigenvectors of (\ref{ci29}) are thus  related as
\begin{equation}
 A_1=Y_0\, \overline A_0,
\end{equation}
where the complex parameter $Y_0$ comprises
\begin{equation}\label{ci32}
 Y_0=\frac{1}{\overline U_1}\left[-E_1\pm \sqrt{\frac{\sqrt{\alpha^2+\beta^2}-\alpha}{2}}-i\left(E_0+V_1\pm\sqrt{\frac{\sqrt{\alpha^2+\beta^2}+\alpha}{2}}\right)\right].
\end{equation}
Therefore, the quaternionic amplitude (\ref{ci35}) conforms to
\begin{equation}\label{ci36}
 \mathcal A=\big(1+Y_0\, j\big)A_0.
\end{equation}
One can understand the $Y_0$ parameter as a measure of the deviation of the quaternionic wave function from the complex wave function, and also as the degree of interaction between the complex components of the wave function. Interesting particular cases require
\begin{equation}\label{ci33}
 E_0+V_1=0,\qquad \mbox{and}\qquad E_1^2>V_0^2+U_1\overline U_1,
\end{equation}
and a pure stationary state on the spatial coordinate assumes
\begin{equation}\label{ci39}
K_0^2=0,\qquad \mbox{and}\qquad K_1^2=\frac{2m}{\hbar^2}\Big(\sqrt{E_1^2-U_1\overline U_1}-V_0\Big).
\end{equation}
Inversely, the pure non-stationary spatial wave function  parameters are
\begin{equation}
 K_0^2=\frac{2m}{\hbar^2}\Big(V_0-\sqrt{E_1^2-U_1\overline U_1}\Big), \qquad \mbox{and}\qquad K_1^2=0.
\end{equation}
For both of these cases, using (\ref{ci32} -\ref{ci33}) one obtains the complex parameter $Y_0$ to be the complex quantity
\begin{equation}
 Y_0=\frac{E_1}{\overline U_1}\left(\pm\sqrt{1-\frac{U_1\overline U_1}{E_1^2}}-1\right).
\end{equation}
These results finally authorize one to consider the self-interacting particle confined within the infinitely deep cavity.

\paragraph{QUATERNIONIC INFINITELY DEEP CAVITY} In analogy to (\ref{ci05}), the quaternionic potential reads
\begin{equation}
\mathcal U(x)=\left\{
\begin{array}{ll}
U      & \qquad \mbox{if}\qquad x\in\left[-\nicefrac{\ell}{2},\,\nicefrac{\ell}{2}\right] \\ \\
\infty & \qquad \mbox{if} \qquad x\in\left(-\infty,\,-\nicefrac{\ell}{2}\right)\cup \left(\nicefrac{\ell}{2},\,\infty\right).
\end{array}
\right.
\end{equation} 
Because of (\ref{ci36}) the quaternionic wave function of the autonomous particle (\ref{ci23}) becomes 
\begin{equation}
 \Psi= \big(1+Y_0\, j\big)A\exp\left[K x-\frac{E}{\hbar}t\right],
\end{equation}
where $A$ is complex. Therefore, the quaternionic wave function within an infinitely deep cavity is also is remarkably similar to the complex case (\ref{ci02}-\ref{ci34}), namely
\begin{equation}\label{ci15}
 \Psi(x,\,t)=\Phi\big(x\big)\exp\left[-\frac{E}{\hbar}t\right]
\end{equation}
whereby the energy parameter $E$ is complex. The space dependent function thus comprehends
\begin{equation}
 \Phi(x)=\big(1+Y_0\, j\big)\Big(A\exp\left[Kx\right]+B\exp\left[-Kx\right]\Big).
\end{equation}
Therefore, the boundary conditions (\ref{ci06}-\ref{ci07}) of the complex case hold identically to the quaternionic case. Consequently, the complex wave functions (\ref{ci37}), (\ref{ci16}), (\ref{ci24}, (\ref{ci25}), and (\ref{ci38}), all of them turn quaternionic through the simple multiplicative unitary quaternionic factor
\begin{equation}\label{ci30}
 \Psi=\frac{1+Y_0j}{\sqrt{1+Y_0\overline Y_0\,}}\psi.
\end{equation}
Moreover, the expectation values of the position, and of the linear momentum will also not change, therefore being absolutely identical to the complex case. One could thus suppose the situations to be physically equivalent, and therefore the quaternionic cavity would be unable of physical novelties. However, the energy of the quaternionic cases presents a clear originality. Remembering 
(\ref{ci39}), the quantization condition give
\begin{equation}\label{ci45}
 E_1(N)=\sqrt{E_N^2+U_1\overline U_1},
\end{equation}
where
\begin{equation}\label{ci41}
 E_N=\frac{\hbar^2\pi^2N^2}{2m\ell^2}+V_0
\end{equation}
is the energy parameter of the complex case, which is recovered within the limit $U_1=0$,  as expected. However,  it is not possible to identify a quantum of energy which every quaternionic state encompasses a multiple of. In other words, every complex energy is multiple of a quantum of energy, and consequently the energetic difference between complex is,
\begin{equation}
 E_N-E_M= \Big(N^2-M^2\Big)\frac{\hbar^2\pi^2N^2}{2m\ell^2}.
\end{equation}
Inversely, the quaternionic states are such that
\begin{equation}
 E_1(N)-E_1(M)=\sqrt{E_N^2+U_1\overline U_1}-\sqrt{E_M^2+U_1\overline U_1},
\end{equation}
and the quaternionc case approximates the complex situation either in case of very high quantum numbers $N$ and $M$, or very small self-interaction $U_1\overline U_1$. However, the difference of the squared energies is an integer multiple of the squared quantum of the complex energy,
\begin{equation}
 E_1^2(N)-E_1^2(M)=\Big(N^4-M^4\Big)\left(\frac{\hbar^2\pi^2N^2}{2m\ell^2}\right)^2
\end{equation}
an interesting feature to be further investigated in future research.

In summary, one can say that the self interaction does not modify the features of the problem that are dependent on their symmetry, like the expectation values of the position and momentum of the particle. However, it changes remarkably the structure of the energy levels. The quaternionic case is thus a clear generalization of the complex case enabled with a richer structure, as expected.

\section{QUATERNIONIC CAVITY II}
Alternatively to (\ref{ci21}), the wave equation
\begin{equation}\label{ci42}
 \hbar\frac{\partial\Psi}{\partial t}i=\widehat{\mathcal H}\Psi,
\end{equation}
generates a system of two complex equations together with the wave function (\ref{ci20}), so that
\begin{eqnarray}\nonumber 
&&\;\;\; i\hbar\frac{\partial\psi_0}{\partial t}=-\frac{\hbar^2}{2m}\nabla^2\psi_0+U_0\psi_0-U_1\psi_1^\dagger\\
&& - i\hbar\frac{\partial\psi_1}{\partial t}=-\frac{\hbar^2}{2m}\nabla^2\psi_1+U_0\psi_1+U_1\psi_0^\dagger.
\end{eqnarray}
The autonomous particle solution, early considered in \cite{Giardino:2024tvp}, deploys constant scalar potentials $U_0$ and $U_1$ and produces the matrix equation
\begin{equation}
\left[
 \begin{array}{cc}
  U_0+i E & -\, U_1\\
  \overline{ U}_1 & \overline U_0+i\, E
  \end{array}
\right]\left[
\begin{array}{c}
 A_0 \\
 \overline A_1
\end{array}
\right]
=
\frac{\hbar^2K^2}{2m}
\left[
\begin{array}{c}
 A_0 \\
 \overline A_1
\end{array}
\right].
\end{equation}
Accordingly, the solution will be
\begin{eqnarray}
&&\label{ci47} K_0^2=\frac{m}{\hbar^2}\left[V_0-E_1+\sqrt{\big(V_0-E_1\big)^2+\left(E_0\pm\sqrt{V_1^2+U_1\overline U_1}\right)^2}\,\right]\\
&&\label{ci48} K_1^2=\frac{m}{\hbar^2}\left[E_1-V_0+\sqrt{\big(V_0-E_1\big)^2+\left(E_0\pm\sqrt{V_1^2+U_1\overline U_1}\right)^2}\,\right].
\end{eqnarray}
The wave function thus follows the previous quaternionic solution (\ref{ci15}-\ref{ci30}), and the only difference corresponds to the factor
\begin{equation}
 Y_0=-\frac{i}{\overline U_1}\left[V_1+E_0\pm\left( E_0\pm\sqrt{V_1^2+U_1\overline U_1}\right)\,\right]
\end{equation}
Of course, the choice
\begin{equation}
 E_0=V_1=U_1=0,\qquad \mbox{and}\qquad V_0<E_1,
\end{equation}
whereby $K_0=0$, recovers the non-self-interaction solution developed in  \cite{Giardino:2020cee}, and nothing new emerges. On the other hand, one obtains a stationary process along the spatial coordinate whenever 
\begin{equation}\label{ci40}
 E_0\pm\sqrt{V_1^2+U_1\overline U_1}= 0\qquad \mbox{and}\qquad V_0<E_1.
\end{equation}
Because $K_0=0$, the energy can be quantized, recovering (\ref{ci41}). Nonetheless, $E_0\neq 0$, forces non-stationary behavior within the time coordinate, and a authentic stationary state is not reached, and this is the principal different from the previous quaternionic case, where stationary self-interacting solutions are possible. The interpretation follows the previous quaternionic case, and therefore, it does not add further possibilities to be discussed.

\section{CONCLUSION}

In this article the well known quantum problem of the infinitely deep cavity has been considered using the real Hilbert space approach, originally developed as a framework for quaternionic quantum mechanics ($\mathbbm H$QM). As expected, the results reproduce the well known results of the ordinary complex quantum mechanical ($\mathbbm C$QM), and presents novel features of the complex solutions, particularly concerning stationary states that satisfy the condition $\exp[K\ell]=\pm 1$ and the non-stationary states.

	 In the case of the quaternionic states, the solution of the the totally general cavity was an open problem, and has been solved for the first time. The presence of the self-interaction within the quaternionic states promoted by the quaternionic components $U_1$ of the scalar potential (\ref{ci46}) implies different energy levels from the complex case, as can be seen in (\ref{ci45}), where stationary states are allowed. On the other hand, the parameters (\ref{ci47}-\ref{ci48}) of the second quaternionic solution do not allow simultaneously time and spacial stationary solutions, another difference from the usual complex solutions. Even though the quaternionic results do not match the complex cases, these complex situations are always recovered within certain limits, demonstrating the generalized character of the quaternionic formalism, and also unraveling a novel and richer  structure of energy levels and physical motion.
	 

There are several topics to be considered as directions of future research. The most obvious concern the remaining problems of one-dimensional quantum mechanics in order to evaluate the effect of self-interaction. After this, one expects to apply this formalism in relativistic quantum theory, where a wide range of problems remains to be examined.

\paragraph{DATA AVAILABILITY STATEMENT} The author declares that data sharing is not applicable to this article as no data sets were generated or analysed during the current study.

\paragraph{DECLARATION OF INTEREST STATEMENT} The author declares that he has no known competing financial interests or personal relationships that could have appeared to influence the work reported in this paper.

\paragraph{FUNDING} This work is supported by the Funda\c c\~ao de Amparo \`a Pesquisa do Rio Grande do Sul, FAPERGS, grant 23/2551-0000935-8 within edital 14/2022.

%
%
%
%

\begin{footnotesize}

\end{footnotesize}
\end{document}